\documentclass[10pt,conference]{IEEEtran}
\IEEEoverridecommandlockouts

\usepackage{cite}

\usepackage{amsmath}
\usepackage{verbatim}

\usepackage{esvect}

\usepackage[ruled,linesnumbered]{algorithm2e} 

\usepackage{graphicx}
\graphicspath{ {figure/} }
\usepackage{multirow}
\usepackage{amsfonts,amssymb}
\usepackage{color}
\usepackage{epstopdf}
\usepackage{bigstrut}
\usepackage{caption}
\usepackage{subcaption}

\usepackage{url}
\usepackage{wrapfig}
\usepackage{amssymb}
\usepackage{mathtools}

\usepackage[ruled]{algorithm2e}

 \usepackage{setspace} 
\usepackage{tabularx,booktabs}
\newcolumntype{Y}{>{\centering\arraybackslash}X}
\usepackage{nicefrac}

\let\labelindent\relax
\usepackage{enumitem}
\usepackage{upgreek}

\usepackage{tabularx,booktabs}
\newcolumntype{Y}{>{\centering\arraybackslash}X}

\begin{document}


%





\title{DynaMarks: Defending Against Deep Learning Model Extraction Using Dynamic Watermarking 
\vspace{-8mm}
}


\author{\IEEEauthorblockN{Abhishek Chakraborty, Daniel Xing, Yuntao Liu, and Ankur Srivastava}
\IEEEauthorblockA{Department of Electrical and Computer Engineering, University of Maryland, College Park, MD, USA \\\{abhi1990,dxing97,ytliu,ankurs\}@umd.edu\\
}}

\maketitle


\begin{abstract}
The functionality of a deep learning (DL) model can be stolen via model extraction where an attacker obtains a surrogate model by utilizing the responses from a prediction API of the original model. 
In this work, we propose a novel watermarking technique called {DynaMarks} to protect the intellectual property (IP) of DL models against such model extraction attacks in a black-box setting.
Unlike existing approaches, {DynaMarks} does not alter the training process of the original model but rather embeds watermark into a surrogate model by dynamically changing the output responses from the original model's prediction API based on certain secret parameters at inference runtime.
The experimental outcomes on Fashion MNIST, CIFAR-10, and ImageNet datasets demonstrate the efficacy of {DynaMarks} scheme to watermark surrogate models while preserving the accuracies of the original models deployed in edge devices.
In addition, we also perform experiments to evaluate the robustness of {DynaMarks} against various watermark removal strategies, thus allowing a DL model owner to reliably prove model ownership.
\end{abstract}

\section{Introduction}
\label{sec_wm_intro}



Deep learning (DL) models are increasingly being deployed in a wide variety of real-world applications such as image classification, natural language processing, autonomous vehicles, smart health, automated manufacturing, home assistants, etc.~\cite{goodfellow2016deep}.
The costs associated with the development of DL models for commercial use are typically very high, especially when the network training phase requires (i) collection of massive annotated dataset (ii) allocation of substantial computing resources to tune the underlying model topology, weight parameters, and hyperparameters 
for attaining high accuracy. 
Therefore, a well-trained DL model is considered to be an intellectual property (IP) of the model owner and it needs to be protected against any sort of IP infringement attempts for preserving the owner's competitive edge in business~\cite{rouhani2018deepsigns, jia2020entangled, orekondy2019knockoff, lee2018defending}.
The growing trend of deployment of DL models by major enterprises like Apple, Amazon, Facebook, Google, and Microsoft in their products have greatly expanded the threat surface, allowing attackers to mount {\em model extraction} attacks~\cite{jia2020entangled, szyller2019dawn, lee2018defending, lukas2019deep}.
In a typical model extraction attack, an adversary queries the original model (also referred to as the {\em victim model}) with inputs of her choice (via a prediction API) and uses the prediction responses to label a {substitute dataset}.
Subsequently, the attacker uses this substitute dataset to train a {\em surrogate model} that replicates the functionality of the victim model~\cite{tramer2016stealing, papernot2017practical, pal2019framework}.
The attacker can publicly deploy such extracted models to offer competitive services, thus depriving the model owners of their business advantage.
Hence, protection of the IP embodied in a high quality model is strongly required to sustain the emerging business of DL based commercial applications which require substantial amounts of time, money, and effort to develop.

In the recent past, several digital watermarking approaches for deep neural networks (DNNs) have been proposed for defending against model IP theft~\cite{adi2018turning, merrer2017adversarial, rouhani2018deepsigns, uchida2017embedding, guo2018watermarking,zhang2018protecting,nagai2018digital}.
Such schemes can be broadly classified into two categories
(i) {\em white-box watermarking} where a DL model owner requires access to the parameters of a suspected stolen model for verifying the embedded watermark
(ii) {\em black-box watermarking} where the watermark can be successfully detected by studying the responses (output probabilities) from the suspected model's prediction API for selected input queries.
In practice, the model parameters as well as the network architecture are usually kept private in commercial services~\cite{kariyappa2020defending} and hence, in this work we focus on developing an effective black-box watermarking scheme to protect the IP rights of DL model owners under more realistic circumstances.
Existing black-box watermarking techniques~\cite{adi2018turning, rouhani2018deepsigns, zhang2018protecting} involve introduction of a set of outlier inputs called the {\em trigger set} with incorrectly assigned labels (known only to the DL model owner) in the training data. 
Subsequently, if the model owner suspects a model to be a stolen copy of his model then he can claim ownership by demonstrating the knowledge of the prediction outcomes for the trigger set.
However, watermarking based on such owner-selected trigger set is {\em ineffective} against model extraction attack as in this case it is the attacker who controls the training set of a surrogate model~\cite{szyller2019dawn}.
As long as she queries a watermarked victim model using inputs sampled from the task distribution, a derived surrogate model will retain no information relevant to the watermark (out-of-distribution input-output pairs) on its learned decision boundaries.

In~\cite{jia2020entangled}, the authors propose entangled watermarks to defend against model extraction attack. It alters the training process in a manner such that the model is encouraged to learn features common to both (i) data that is sampled from the task distribution and (ii) data that encodes the watermark.
However, the computational overhead associated with the modified training phase is substantially high and also, the accuracy of a watermarked model is sensitive to hyperparameter tuning.
In~\cite{szyller2019dawn}, another watermarking scheme called DAWN has been proposed to counter model extraction attack. DAWN does not alter the training process of a DL model, but instead selectively changes the responses of the model's prediction API to embed watermark.
Although DAWN offers a promising approach to deter model extraction, its applicability is limited to a client-server model where a malicious client (attacker) submits queries to a DL service hosted by the model owner using a trusted server.
In this work, we propose a novel accuracy-preserving DNN watermarking technique called DynaMarks which provides an effective IP security solution for DL models deployed in servers as well as in edge devices against model extraction attacks.
Unlike existing approaches, the proposed scheme doesn't introduce any computational overhead in the model training process nor does it require
the support of a trusted server to watermark surrogate models.
DynaMarks generates watermark by dynamically altering the responses (output probabilities) of a DL model's prediction API based on certain secret parameters at inference runtime.
Any surrogate model trained using such responses will retain the watermark information pertaining to the model owner's secret parameters.
We summarize the contribution of this paper as follows:
\vspace{-1.7mm}
\begin{itemize}
    \item Proposing DynaMarks, a black-box watermarking technique to defend against model extraction attacks on proprietary DL models deployed in edge devices. DynaMarks embeds watermark by dynamically changing the responses of the model's prediction API during the inference phase using low-cost hardware, without introducing any computational overhead in the training process. 
    \item Performing extensive evaluations using different DNN architectures and datasets (Fashion MNIST, CIFAR-10, and ImageNet) to demonstrate the effectiveness of DynaMarks to watermark stolen surrogate models. The embedded watermark is detected by comparing the output probability distributions of the victim and surrogate models for a given set of queries.
    \item Empirically assessing the robustness of DynaMarks against different types of watermark removal attacks. Our experimental findings reveal that stolen surrogate models retain watermarks even when subjected to various model transformations, thus highlighting the efficacy of the proposed DynaMarks scheme. 
\end{itemize}

\section{Background}


\subsection{Model Extraction Attacks}
A DNN classifier is a function $\mathcal{F}: \mathbb{R}^M\rightarrow\mathbb{R}^N$, where $M$ and $N$ are the number of input features and output classes respectively.
The output of $\mathcal{F}(x)$ on input $x$ is an $N$-dimensional vector $\vv{p_x}$ containing probabilities $p^j_x$ that $x$ belongs to class $c_j$ for $j \in [N]$\footnote[1]{$[N]$ denotes the set of first $N$ natural numbers.}.
The predicted class $\mathcal{C}$ corresponds to the output component with maximum value as obtained by applying {\em argmax} function: ${\mathcal{C}}$ = $argmax \ \mathcal{F}(x)$.
In practice, the DNN classifier is trained using a massive annotated dataset along with an optimized set of hyperparameters such that $argmax \ \mathcal{F}(x)$ approximates the  oracle function $\mathcal{O}$ which outputs the true class label for any input sample $x \in \mathbb{R}^M$.

Several model extraction attacks~\cite{orekondy2019knockoff,atli2019extraction} against machine learning models (including complex DNNs) have been proposed in recent literature which pose a major threat to the IP rights of their owners.
In a model extraction attack, the objective of an attacker is to {steal} the functionality of a well-trained network $\mathcal{F}_{org}$ (referred to as the original or victim model) by querying it with a set $\mathbb{Q}$ of input queries and obtaining the corresponding set of predicted output probabilities $\mathcal{F}_{org}(\mathbb{Q})$. 
The attacker uses this information to train a surrogate model $\mathcal{F}_{sm}$ such that its accuracy is close to that of $\mathcal{F}_{org}$ on a test dataset, thus depriving the DL model owner of his business advantage. 
Typically, a model extraction attack is performed in black-box setting, i.e., the attacker doesn't have any knowledge of the weight parameters of $\mathcal{F}_{org}$, but has access to its prediction API which returns the output probabilities for a given input query.
The challenges associated with such an attack strategy are 
(i) lack of availability of annotated training data that comes from the {same} distribution as the data used to train $\mathcal{F}_{org}$
and (ii) no knowledge of the architecture of $\mathcal{F}_{org}$ or its training process.
In this work, we focus on watermarking based approaches to defend against model extraction attacks on proprietary DL models.


\subsection{Black-box DNN watermarking}
\label{sec:background_dnn_wm}

Digital watermarking is a popular technique utilized to covertly embed a secret marker into the cover data such as images, videos, or audios. It enables free sharing of digital content, while providing proof of ownership of the cover data.
Extension of watermarking approaches to deep learning offers an effective solution to defend against model theft by allowing the owner to claim IP rights upon inspection of a suspected stolen model.
Several existing black-box DNN watermarking approaches~\cite{adi2018turning, rouhani2018deepsigns, zhang2018protecting} consist of {\em overfitting} a model $\mathcal{F}_{org}$ to outlier input-output pairs (known only to the DL model owner).
Such watermarking techniques are based on the concept of backdoor insertion~\cite{adi2018turning, gu2017badnets} using a {trigger set}.
If the DL model owner encounters a model which exhibits targeted misclassifications on this trigger set that was encoded by the watermark, then the owner can reasonably claim that the model is a stolen copy of $\mathcal{F}_{org}$.

\noindent{\bf Entangled Watermarks.} In~\cite{jia2020entangled}, the authors demonstrated a fundamental limitation of conventional DNN watermarking schemes based on outlier input-output pairs in the context of model extraction attack.
In order to perform model extraction, an attacker does not directly steal the original model, but rather trains a surrogate model by using the information obtained by querying the original model.
If the attacker queries a watermarked model $\mathcal{F}_{org}$ using inputs which are sampled from the task distribution, then the obtained surrogate model $\mathcal{F}_{sm}$ will only learn the victim model's decision surface relevant to the task distribution and will not retain the decision surface relevant to watermarking.
Subsequently, the authors propose a new technique called entangled watermarks which trains a DL model to learn features common to both task distribution and watermark data by formulating a new loss function. 
However, this altered training process incurs a substantial increase (about 2$\times$ compared to baseline model) in computational overhead.
Moreover, the success of entangled watermarks scheme is heavily dependent on the training dataset as well as on hyperparameter tuning. In fact, such a watermarked model will incur severe performance degradation if the hyperparameters are not carefully selected, leading to a decrease in model utility. 

\noindent{\bf DAWN.} In~\cite{szyller2019dawn}, the authors propose a technique called DAWN which does not impose any alterations to the training process but selectively changes the responses of a model's prediction API in order to watermark a fraction of input queries.
These watermarked queries then serve as a trigger set for a surrogate model $\mathcal{F}_{sm}$ trained using the API responses of the victim model $\mathcal{F}_{org}$.
Unlike prior  backdoor insertion based watermarking schemes, DAWN is resilient to model extraction attacks as all the input queries including those which are watermarked belong to the task distribution, i.e., no outlier inputs are present in the trigger set.
Although an effective technique to deter model extraction, the effectiveness of DAWN is limited to client-server model where a malicious client (attacker) submits queries to $\mathcal{F}_{org}$ hosted by the DL model owner using a trusted server.
Such a scheme will not be applicable in scenarios where the model owner has no knowledge of the input queries made by the attacker, e.g., DL models deployed in remote edge devices.
Also, returning false predictions with the objective of embedding watermarks can be unacceptable for certain applications, e.g., malware detection, medical applications such as cancer diagnosis, etc.~\cite{atli2019extraction, wang2019robust}.
Moreover, DAWN does not secure against model extraction attack which utilizes several similar queries (multiple close images) where only one is assigned a false label~\cite{lukas2019deep}.
This strongly motivates the need to develop an effective watermarking scheme which addresses the above drawbacks.


\section{Problem Description}
\label{sec:DynaMarks_ProblemDescription}

In recent years, there is a growing trend toward transition of the inference phase of deep learning to edge devices~\cite{li2018edge, wu2019machine}. This leads to improved user experience with reduction in inference time (low latency), less dependency on network connectivity, and increased energy efficiency of resource-constrained mobile devices.
In addition, running inference on the edge also enables several deep learning services, e.g. Instagram features that involve real-time application of machine learning algorithms at image capture time~\cite{wu2019machine}.
In this new paradigm of edge intelligence, an attacker can directly query a proprietary DL model deployed in an edge device without any need to redirect the queries to a trusted cloud server, thus rendering model protection countermeasures like DAWN~\cite{szyller2019dawn} practically useless. 
Even the utilization of a trusted hardware in the edge as proposed in~\cite{chakraborty2020hardware} will not be effective for 
protecting DL model IPs as any end-user possessing an authorized hardware will be able to mount model extraction attack.
Therefore, the emerging trend of executing deep learning inference on edge devices poses major challenges to the security of well-trained DL models from IP infringement attempts.
In this work, we propose a novel accuracy-preserving watermarking approach called DynaMarks as an effective IP protection mechanism 
for DL models deployed in edge devices.


\noindent {\bf Watermarking Requirements.}
The following requirements should be addressed while designing an effective black-box watermarking scheme for proprietary DL models deployed in edge devices:
\vspace{-2mm}
\begin{itemize}[itemsep=1pt,topsep=1pt,leftmargin=10pt]
    \item {\bf Fidelity.} The performance or accuracy of the original DNN classifier should not degrade due to watermark embedding.
    
    \item {\bf Robustness.} The embedded watermark should exhibit resiliency against model modifications such as compression/pruning and provide high detection confidence for proving model ownership.
    
    
    \item {\bf Imperceptibility.} The watermark should not leave tangible footprints in the model, thus hindering any unauthorized detection.
\end{itemize}

\vspace{-2mm}
\noindent The above set of requirements have also been considered in previous works on black-box DNN watermarking~\cite{adi2018turning, rouhani2018deepsigns, zhang2018protecting, merrer2017adversarial}. In addition, we consider the following important watermarking requirement for defending against model extraction attacks.
\vspace{-2mm}
\begin{itemize}[itemsep=1pt,topsep=1pt,leftmargin=10pt]
\item {\bf Transferability.} The embedded watermark should survive model extraction attack, i.e., the watermark should get transferred from the original DL model to a surrogate model obtained from it.
\end{itemize}

\noindent {\bf Threat Model.}
The objective of an attacker is to extract the functionality of a well-trained DL model $\mathcal{F}_{org}$ without its watermark~\cite{jia2020entangled}.
In practice, the attacker is data-limited~\cite{kariyappa2020defending, orekondy2019prediction} and does not have sufficient number of inputs representative of the training set of $\mathcal{F}_{org}$.
In this paper, we assume an adversary with the following capabilities:\\
\noindent{\bf (i)} access to $\gamma$ fraction of the training data of $\mathcal{F}_{org}$ but not its labels, constituting the attacker's input query set $\mathbb{Q}$\\
\noindent{\bf (ii)} knowledge of the network architecture of $\mathcal{F}_{org}$\\
\noindent{\bf (iii)} ability to query $\mathcal{F}_{org}$ with any input sample and obtain the output probability for each class (black-box setting)\\
\noindent{\bf (iv)} knowledge that $\mathcal{F}_{org}$ is watermarked but does not know the details of the watermarking procedure.\\
\hspace*{4mm}In addition, we also assume that $\mathcal{F}_{org}$ is deployed in a remote edge device and the DL model owner does not have any influence on the query strategy or the training process adopted by the attacker to obtain the surrogate model $\mathcal{F}_{sm}$. Furthermore, the attacker is not constrained by memory or computational capabilities.






\vspace{-1mm}
\section{DynaMarks}
\label{sec:DynaMarks}
\vspace{-1mm}

In this section, we present a new black-box watermarking technique called DynaMarks to defend against model extraction attacks on proprietary DL models.
In order to perform model extraction, the attacker queries the original model with inputs from her query set $\mathbb{Q}$ and utilizes the predicted output probabilities to train a surrogate model.
Previous works~\cite{tramer2016stealing, lee2018defending} have shown that using output probabilities instead of labels drastically reduces (about $50$-$100\times$) the number of queries required to extract the model and also, improves the attack convergence as well as increases the converged model accuracy.
The objective of our proposed DynaMarks scheme is to smartly alter these output probabilities in order to watermark a surrogate model without sacrificing the prediction accuracy of the original DL model.




\vspace{-1mm}
\subsection{Watermark Embedding}
\label{sec:DynaMarks_wm_embed}
\vspace{-1mm}

As per our threat model, the primary challenge for DynaMarks technique is to embed watermark into the original model $\mathcal{F}_{org}$ running on an edge device in such a way that during model extraction the watermark gets transferred to the surrogate model $\mathcal{F}_{sm}$.
Since the DL model owner does not have any control over the input queries used by the attacker, she can no longer utilize the notion of a trigger set for embedding watermark as adopted in several prior approaches~\cite{adi2018turning, rouhani2018deepsigns, zhang2018protecting, szyller2019dawn}.
Instead, DynaMarks dynamically alters the output probabilities of $\mathcal{F}_{org}$ based on certain secret parameters at model inference runtime with the objective of watermarking $\mathcal{F}_{sm}$. Next, we present the details of the watermark embedding process.


\begin{algorithm}[t]
 \KwIn{(i) Network $\mathcal{F}_{org}$ initialized with weights $\mathcal{W}_{org}$
 
 (ii) Vectors $\vv{V_i}$ and parameters ($\alpha_i$, $\beta_i$, $a_i$, $b_i$), $\forall i \in [N]$
 
 (iii) Attacker's input query set $\mathbb{Q}$
 
 }
 \KwOut{Watermarked surrogate model $\mathcal{F}_{sm}$ 
 
 
 }\BlankLine

\vspace{1mm}
{\fontfamily{qcr}\selectfont/*building substitute dataset*/}

$D_{sub}$ = $\phi$ 

\For{ $x \in \mathbb{Q}$}{ 

    $\vv{p_x} = \mathcal{F}_{org}(x)$ 

    
            \If { $argmax \ \vv{p_x}$ == $i$ and $ p^i_x \in (\alpha_i, \beta_i) $}{
            
                Generate random variable $\Delta p \thicksim U (a_i, b_i)$
                
                Select an index $j \in N$ from $\vv{V_i}$ 
                
                {\fontfamily{qcr}\selectfont /*alter component pair $(p^i_x,p^j_x)$*/}
                
                $p^i_x$ = $p^i_x - \Delta p$
                
                $p^j_x$ = $p^j_x + \Delta p$
                
                
                
                    
                        
                        
                    
            }

    $D_{sub}$ = $D_{sub} \cup (x,\vv{p_x})$
}

\vspace{2mm}
{\fontfamily{qcr}\selectfont /*watermark transferability*/}

$\mathcal{W}_{sm} \leftarrow$ Train $\mathcal{F}_{sm}$ using $D_{sub}$

\bf{return} $\mathcal{F}_{sm}$

\caption{Watermark Embedding Process on Extracted Model $\mathcal{F}_{sm}$}
\label{algo_wm_embed}
\end{algorithm}

Let us denote the output probability vector of $\mathcal{F}_{org}(x)$ for an input $x$ by $\vv{p_x}$.
Note that each component $p^i_x$ of the vector $\vv{p_x}$ corresponds to the probability value predicted by the model for the $i^{th}$ class, $i \in [N]$.
Now, for every $i^{th}$ class, the DL model owner defines a {\bf secret} vector $\vv{V_i} = (v^1_i, v^2_i, \cdots, v^N_i)$ of length $N$ such that the values of its elements specify the selection probabilities of the corresponding indices, e.g., the probability of selecting the $j^{th}$ index of $\vv{V_i}$ is $v^j_i$.
The process of embedding watermark into the model $\mathcal{F}_{sm}$ by utilizing these secret vectors is summarized in Algorithm~\ref{algo_wm_embed}.
In order to extract the functionality of $\mathcal{F}_{org}$, the attacker uses samples from the input query set $\mathbb{Q}$ to query $\mathcal{F}_{org}$ and builds a substitute dataset $D_{sub}$ with the returned responses (output probabilities). 
For each input $x \in \mathbb{Q}$, DynaMarks alters the output probability vector $\vv{p_x}$ as a function of vectors $\vv{V_i}$ as follows (lines $4$-$11$ of Algorithm~\ref{algo_wm_embed}):
If the $i^{th}$ component of $\vv{p_x}$ has the maximum value, i.e., $argmax \ \vv{p_x}$ = $i$, and the maximum value $p^i_x$ lies within a certain range $(\alpha_i, \beta_i)$, then 
(i) generate a random variable $\Delta p$ that follows a certain distribution, say a uniform distribution $U(a_i,b_i)$ 
(ii) select an index $j$ from the vector $\vv{V_i}$
(iii) alter the probabilities of component pair $(p^i_x,p^j_x)$ of the vector $\vv{p_x}$ by transferring an amount of $\Delta p$ from $p^i_x$ to $p^j_x$, i.e., $p^i_x$ = $p^i_x - \Delta p$ and $p^j_x$ = $p^j_x + \Delta p$.
Note that such alteration to vector $\vv{p_x}$ does not affect the sum of its components which still equals $1$.
In addition to the vectors $\vv{V_i}$, the parameters ($\alpha_i$, $\beta_i$, $a_i$, and $b_i$), $\forall i \in [N]$, are also {\bf secrets} chosen by the DL model owner.
For a given set of inputs, such probabilistic changes in the output responses of $\mathcal{F}_{org}$ based on the secret parameters leads to a set of altered probability distributions over the set of outputs which constitute the watermark in our proposed DynaMarks scheme.
For notational simplicity, we refer to this altered version of $\mathcal{F}_{org}$ as $\mathcal{F}_{alt}$.

When the attacker uses the responses of $\mathcal{F}_{alt}$ to the input query set $\mathbb{Q}$ for composing the substitute dataset $D_{sub}$ and subsequently, trains a surrogate model $\mathcal{F}_{sm}$ using it, we expect the watermark to get transferred to $\mathcal{F}_{sm}$.
This is because the extracted model $\mathcal{F}_{sm}$ tries to replicate the secret-dependent functionality of $\mathcal{F}_{alt}$ which maps a set of inputs to a set of altered probability distributions over the set of outputs.
The above methodology of embedding watermark into $\mathcal{F}_{sm}$ doesn't require the DL model owner to have any knowledge of the attacker's query strategy. Hence, the proposed DynaMarks technique provides an effective IP security solution against model extraction attacks on proprietary DL models deployed in edge devices.
Also, for a given dataset and network architecture, the model owner can choose the secret parameters in such a manner that the accuracy of the original model $\mathcal{F}_{org}$ is preserved.



\begin{algorithm}[t]
 \KwIn{(i) Black-box access to network $\mathcal{F}_{sm}$ 
 
 (ii) White-box access to networks $\mathcal{F}_{org}$ and $\mathcal{F}_{alt}$
 
 (iii) Vectors $\vv{V_i}$ and parameters ($\alpha_i$, $\beta_i$, $a_i$, $b_i$), $\forall i \in [N]$
 
 (iv) DL model owner's verification query set $\mathbb{V}$
 
 
 (v) Watermark detection threshold $\tau$
 
 }
 \KwOut{Decision of watermark detection in $\mathcal{F}_{sm}$
 
 }\BlankLine

{\fontfamily{qcr}\selectfont /*Form $N \times N$ response distributions*/}

Initialize response matrices $R^{sm}_{ij}$, $R^{org}_{ij}$, $R^{alt}_{ij}$,   \ \ $\forall i, j \in [N]$


\vspace{0.5mm}
\For{ $(x,y) \in \mathbb{V}$}{
    \vspace{0.5mm}
    $\Vec{s_x} = \mathcal{F}_{sm}(x)$, $\Vec{p_x} = \mathcal{F}_{org}(x)$, $\Vec{q_x} = \mathcal{F}_{alt}(x)$
    \vspace{0.5mm}

    \For{ $j \in [N]$}{ 
    
        Append $s^j_x$, $p^j_x$, $q^j_x$ to $R^{sm}_{yj}$, $R^{org}_{yj}$, $R^{alt}_{yj}$

    }
}

\hspace*{-1.5mm} $D^{sm}_{ij}$, $D^{org}_{ij}$, $D^{alt}_{ij}\hspace*{-0.5mm}  \leftarrow$ Create distributions of $R^{sm}_{ij}$, $R^{org}_{ij}$, $R^{alt}_{ij}$  





\vspace{2.0mm}
{\fontfamily{qcr}\selectfont /*Calculate distance metrics*/}
\vspace{0.8mm}

$\delta^{org}_{sm} \leftarrow 0$, $\delta^{alt}_{sm} \leftarrow 0$

\vspace{0.8mm}
\For{ $i \in [N]$}{
    
    \For{ $j \in [N]$}{
        \vspace{1mm}
        $\delta^{org}_{sm} = \delta^{org}_{sm}$ + JSD$(D^{org}_{ij} \parallel D^{sm}_{ij})$
    
        \vspace{1.3mm}
        $\delta^{alt}_{sm} \hspace{0.5mm} = \delta^{alt}_{sm}$ \hspace{0.5mm}+ JSD$(D^{alt}_{ij} \hspace{0.7mm}\parallel D^{sm}_{ij})$
    
    }

}

\vspace{1.0mm}
{\fontfamily{qcr}\selectfont /*Determine if model is watermarked*/}
\vspace{0.8mm}


$\eta$ = $\delta^{org}_{sm}$ / $\delta^{alt}_{sm}$
\vspace{0.5mm}

\If { $\eta > \tau$ }{
    \vspace{1mm}
    {\bf return} Watermark Detected in $\mathcal{F}_{sm}$
    \vspace{0.8mm}
}


\caption{Watermark Verification Process}
\label{algo_wm_verify}
\end{algorithm}

\subsection{Watermark Verification}

In order to verify the presence of watermark in $\mathcal{F}_{sm}$, the DL model owner compares the distributions of output responses of the models $\mathcal{F}_{sm}$ and $\mathcal{F}_{alt}$ using a verification query set $\mathbb{V}$.
An element of the set $\mathbb{V}$ consists of a tuple $(x,y)$, where $x$ denotes an input query and $y$ denotes its known output label.
In our experiments (details later in section~\ref{sec:DynaMarks_evaluations}),
we consider the entire test set of a benchmark dataset as the verification query set $\mathbb{V}$.
If the model $\mathcal{F}_{sm}$ is stolen from $\mathcal{F}_{alt}$ using model extraction, we expect the distributions of their output responses to be {\em similar} as both of them will be functions of the secret parameters used to embed the watermark.
The process of watermark detection in a suspected surrogate model $\mathcal{F}_{sm}$ is summarized in Algorithm~\ref{algo_wm_verify}.
In the first phase (lines $1$-$9$), the DL model owner defines a data structure called {\em response matrix} $R^{sm}$ (corresponding to model $\mathcal{F}_{sm}$) of dimension $N \times N$ whose each element $R^{sm}_{ij}, \forall i, j \in [N]$, is a {variable-length} list.
The process of populating the elements of $R^{sm}$ is as follows:
The model $\mathcal{F}_{sm}$ is queried with an input $x$ from the set $\mathbb{V}$ to obtain an output probability vector $\Vec{s_x}$ in black-box setting.
If the actual label corresponding to input $x$ is $y$, then the components $s^j_x$ of $\Vec{s_x}$, $\forall j \in [N]$, are appended to the respective lists $R^{sm}_{yj}$ along the $y^{th}$ row of matrix $R^{sm}$.
The repetition of this step for all the elements of set $\mathbb{V}$ results in clustering of the output responses of the model $\mathcal{F}_{sm}$ according to their actual class labels along the rows of the matrix $R^{sm}$.
Subsequently, the DL model owner forms a {\em response distribution} $D^{sm}$ of dimension $N \times N$ using the response matrix $R^{sm}$ by creating individual probability distributions $D^{sm}_{ij}$ from the contents of the corresponding list $R^{sm}_{ij}$, $\forall i, j \in [N]$.
Similarly, the response distributions $D^{org}$ and $D^{alt}$ are also formed by querying the models $\mathcal{F}_{org}$ and $\mathcal{F}_{alt}$ respectively 
using the {\em same} verification set $\mathbb{V}$.
Note that the model owner has white-box access to both $\mathcal{F}_{org}$ and its altered version $\mathcal{F}_{alt}$ for constructing their respective response distributions.

\begin{table*}[t]
\begin{center}
\vspace*{-0mm}
\resizebox{0.98\textwidth}{!}{
 \begin{tabularx}{\textwidth}{|c||c|c||Y|Y|Y||Y|Y|Y||Y|Y|Y|} 
 \hline
  \multirow{2}{*}{\bf Dataset} & \multicolumn{2}{c||}{\bf Fidelity (accuracy) } & \multicolumn{3}{c||}{\bf Averaging Attack} & \multicolumn{3}{c||}{\bf Pruning ($\kappa=10\%$)} & \multicolumn{3}{c|}{\bf Different Architectures}
 \\ [0.5ex] 
 \cline{2-12}
& $\mathcal{F}_{org}$ &  $\mathcal{F}_{alt}$  &
$\delta^{org}_{sm}$ & $\delta^{alt}_{sm}$ & $\mathbf{\eta}$ &
$\delta^{org}_{sm}$ & $\delta^{alt}_{sm}$  & $\mathbf{\eta}$ &
$\delta^{org}_{sm}$ & $\delta^{alt}_{sm}$  & $\mathbf{\eta}$  \\
 \hline\hline
 Fashion MNIST &  $90.72\%$ & $90.72\%$ & $45.32$ & $17.99$ & $\mathbf{2.52}$ & $41.83$ & $16.03$ & $\mathbf{2.61}$  & $73.72$ & $28.71$ & $\mathbf{2.57}$\\ 
 \hline
 CIFAR-10      &  $84.98\%$ & $84.98\%$ & $41.07$ & $16.02$ & $\mathbf{2.56}$  & $33.25$ & $14.63$ & $\mathbf{2.27}$  & $34.56$ & $17.94$ & $\mathbf{1.93}$\\
 \hline
 ImageNet  & $92.40\%$ & $92.40\%$ & $155.65$ & $86.29$ & $\mathbf{1.80}$ & $121.09$ & $76.72$ &  $\mathbf{1.58}$ & $ 180.33$ & $109.64$ & $\mathbf{1.64}$\\
 \hline
\end{tabularx}
}
\vspace*{-0.2mm}
\caption{Evaluating the fidelity and robustness properties of DynaMarks scheme.}
\label{table:DynaMarks_evaluation}
\vspace*{-7mm}
\end{center}
\end{table*}

In the second phase (lines $10$-$17$), the DL model owner calculates a distance metric $\delta^{org}_{sm}$ by iteratively adding the Jensen-Shannon divergence\footnote{Jensen-Shannon divergence JSD$(P \parallel Q)$ is a metric for measuring the similarity between two probability distributions $P$ and $Q$.} JSD$(D^{org}_{ij} \parallel D^{sm}_{ij})$ between the probability distributions $D^{org}_{ij}$ and $D^{sm}_{ij}$, $\forall i, j \in [N]$.
Similarly, another distance metric $\delta^{alt}_{sm}$ is also obtained by considering the response distributions $D^{alt}$ and $D^{sm}$.
In the final phase (lines $18$-$22$), the DL model owner calculates a parameter $\eta$ which is the ratio of $\delta^{org}_{sm}$ to $\delta^{alt}_{sm}$.
If the value of the parameter $\eta$ is greater than a certain threshold $\tau$ (empirically determined after experimental evaluations), then the presence of watermark is detected in the model $\mathcal{F}_{sm}$.
The rationale behind such a decision being that the response distribution $D^{sm}$ will be more {\em similar} to $D^{alt}$ (smaller $\delta^{alt}_{sm}$) than compared to $D^{org}$ (larger $\delta^{org}_{sm}$) if $\mathcal{F}_{sm}$ is trained using the output responses of $\mathcal{F}_{alt}$.
This will lead to a large value of parameter $\eta$ (greater than $\tau$) implying that $\mathcal{F}_{sm}$ is indeed extracted from $\mathcal{F}_{alt}$.
As evident from the above discussion, the success of such watermark detection in DynaMarks scheme strongly depends on how effectively the dynamically generated watermark gets transferred from the output responses of $\mathcal{F}_{alt}$ to the output responses of $\mathcal{F}_{sm}$ due to model extraction.

\vspace{-0mm}
\noindent{\bf Implementation efficiency.}
The proposed DynaMarks technique can be implemented in edge devices using low-cost hardware components with negligible impact on the performance of deployed DL models.
The alterations in the output probabilities of a model can be easily performed using hardware implementations of pseudo-random number generators along with standard digital adder/subtractor circuits.
The incorporation of such simple hardware designs will have insignificant effects on the model inference time as well as on the energy-efficiency of an edge device.
The secret parameters used in the DynaMarks scheme can be stored securely in a tamper-proof chip such as Trusted Platform Module
(TPM)~\cite{cite_tpm} embedded into the edge device.

\vspace{-1mm}
\section{Evaluations}
\label{sec:DynaMarks_evaluations}

\subsection{Experimental Setup}
\label{sec:DynaMarks_exp_setup}

\noindent{\bf Datasets.} 
We evaluate DynaMarks technique on three popular image datasets,\hspace{1mm}Fashion MNIST~\cite{fashion_mnist_cite}, CIFAR-10~\cite{cifar10_cite}, and ImageNet~\cite{howard2020fastai},
using PyTorch $1.7$ framework.
The classification tasks on these datasets are much more complex compared to the classic MNIST dataset, e.g. models achieving $99\%$ accuracy on MNIST only attain about $90\%$ accuracy on Fashion MNIST dataset.
Fashion MNIST dataset consists of $70,000$ samples (training set of $60,000$ examples and test set of $10,000$ examples) of $28\times 28$ grayscale images, 
whereas CIFAR-10 dataset consists of $60,000$ samples (training set of $50,000$ examples and test set of $10,000$ examples) of $32\times32$ colour images.
For both these datasets, an image is associated with a label from $10$ classes ($N$=$10$).
Also, in case of ImageNet dataset, for simplicity in our experiments we considered a subset of $N$=$10$ classes as outlined in \cite{howard2020fastai}. This simplified ImageNet dataset consists of $13,394$ samples (training set of $12,894$ examples and test set of $500$ examples) of RGB colour images cropped to $224\times 224$ pixels. 

\noindent{\bf Network Architectures.}
In order to obtain the well-trained model $\mathcal{F}_{org}$,
we use Convolutional Neural Networks (CNNs) for Fashion MNIST and CIFAR-10 datasets.
In case of Fashion MNIST dataset, the architecture is composed of $2$ convolutional layers with $16$ and $32$ number of $5\times 5$ kernels respectively with $2\times 2$ max pooling operations, followed by a fully-connected layer.
For CIFAR-10 dataset, the architecture is composed of $6$ convolution layers with $32$, $64$, $128$, $128$, $256$, and $256$ $3\times 3$ kernels respectively with $2\times 2$ max pooling operations, followed by three fully-connected layers.
In case of ImageNet dataset, we use the standard ResNet-18 architecture~\cite{he2016deep} to obtain $\mathcal{F}_{org}$.
Each network was trained to attain satisfactory performance on the image classification tasks considered.

\subsection{Validating DynaMarks}

Next, we perform experiments to evaluate the effectiveness of our proposed DynaMarks approach in the context of the watermarking requirements stated in section~\ref{sec:DynaMarks_ProblemDescription}.


\vspace{-1mm}
\noindent{\bf Impact on accuracy (Fidelity requirement).}
The accuracy of a well-trained model $\mathcal{F}_{org}$ should not degrade due to watermark embedding as otherwise its utility will be impacted~\cite{lee2018defending}.
In DynaMarks scheme, a DL model owner has the flexibility to calibrate the secret parameters (vectors $\vv{V_i}$ along with $\alpha_i$, $\beta_i$, $a_i$, and $b_i$, $\forall i \in [N]$) such that the accuracy of the original DNN classifier is preserved.
In our experiments, we construct each $\vv{V_i}$ by setting the
selection probability of a single index (randomly selected from $[N]$) to $2/11$ and the selection probabilities of the remaining indices to $1/11$.
Also, we set $\alpha_i$ = $0.9$, $\beta_i$ = $1$, $a_i$ = $0.01$, and $b_i$ = $0.19$, $\forall i \in [N]$, for all the datasets such that $\mathcal{F}_{alt}$ exhibits accuracy-preserving outcomes as reported in the second and third subcolumns of Table~\ref{table:DynaMarks_evaluation}.


\begin{figure}[!t]
\centering
\vspace*{-5.0mm}
\hspace*{-6mm}\includegraphics[width=1.2\columnwidth, trim = 7cm 0.0cm 5.0cm 0.0cm, clip]{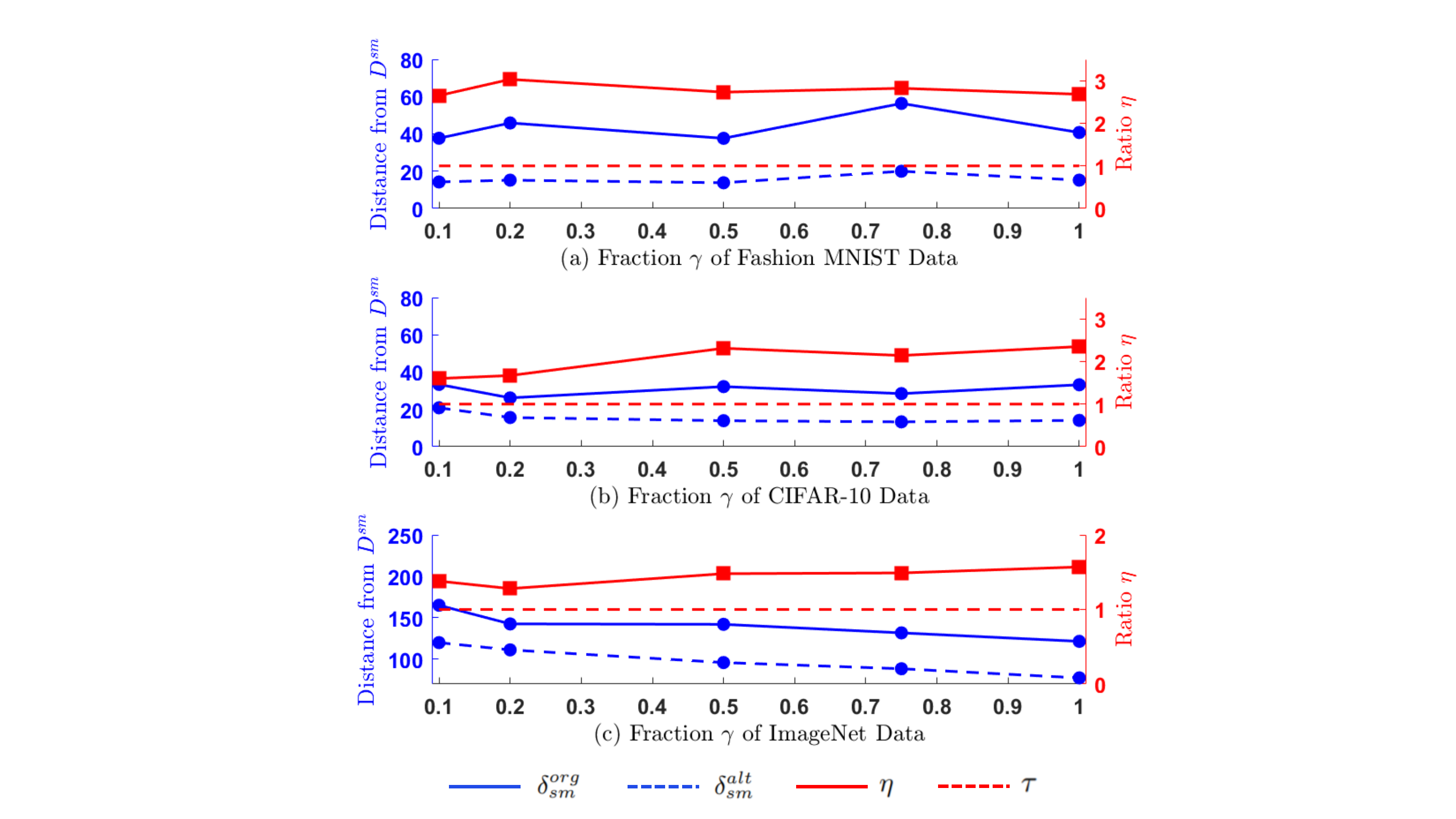}
\vspace*{-3.5mm}
\caption{Transferability of DynaMarks scheme across different training dataset fractions $\gamma$.}
\label{fig_transfer_training_frac}
\end{figure}

\begin{figure}[!t]
\centering
\vspace*{-4mm}
\hspace*{-12mm}\includegraphics[width=1.25\columnwidth, trim = 5.5cm 0.0cm 5.0cm 0.0cm, clip]{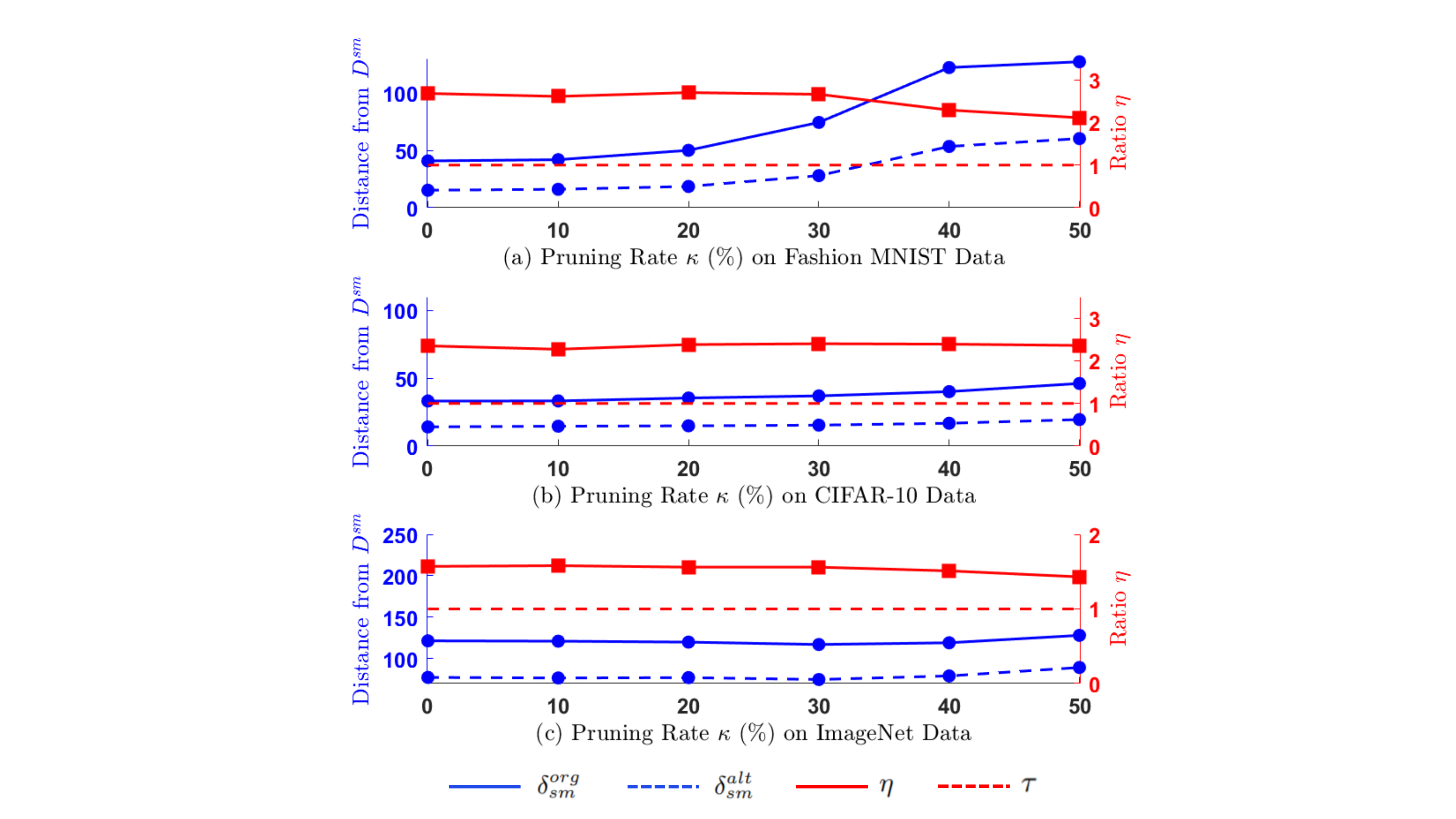}
\vspace*{-3.3mm}
\caption{Transferability of DynaMarks scheme across different model pruning rates $\kappa(\%)$.}
\label{fig_impact_pruning_rate}
\end{figure}

\vspace{-1mm}
\noindent{\bf Detectability in surrogate model (Transferability requirement).} 
A very important criteria for designing watermarking schemes for proprietary DL models deployed in edge devices is the property of watermark transferability from the well-trained model $\mathcal{F}_{alt}$ to a surrogate model $\mathcal{F}_{sm}$ which is extracted from the former.
In fact, the watermark verification process of the DynaMarks scheme (as outlined in Algorithm~\ref{algo_wm_verify}) is also strongly dependent on the detectability of the watermark at the output responses of $\mathcal{F}_{sm}$.
In Fig.~\ref{fig_transfer_training_frac}, we study the transferability property of DynaMarks by evaluating the parameter $\eta$ (shown using solid red line) across different fractions $\gamma$ of the training data available to the attacker.
For all the three datasets, we used the {\em same} network architecture for both $\mathcal{F}_{alt}$ and $\mathcal{F}_{sm}$ and considered the entire test set as the verification set $\mathbb{V}$.
In the subfigures, we also plot the variations of the distance metrics $\delta^{org}_{sm}$ and $\delta^{alt}_{sm}$ using solid and dashed blue lines respectively.
We observe that the parameter $\eta$ is always greater than the threshold $\tau$ = $1$ (no false negatives), implying that there is strong watermark transferability from the output responses of $\mathcal{F}_{alt}$ to the output responses of $\mathcal{F}_{sm}$ due to model extraction.
In order to assess the false positive rate in our watermark verification process, we trained a benign model $\mathcal{F}_{bm}$ using $50\%$ of the original training dataset and recalculated the parameter $\eta$ by taking the ratio of the distances of the output response distribution of $\mathcal{F}_{bm}$ from those of $\mathcal{F}_{org}$ and $\mathcal{F}_{alt}$. In this case, we found that $\eta$ is always lesser than the threshold $\tau$ = $1$ ($\eta$ = $0.19$ for Fashion MNIST, $\eta$ = $0.54$ for CIFAR-10, and $\eta$ = $0.58$ for ImageNet), implying that there is no false positives in the watermark detection process.
The value of the threshold $\tau$ was chosen empirically after performing experimental evaluations on the datasets.
In summary, DynaMarks exhibits strong watermark transferability as required from an ideal watermarking scheme to defend against model extraction attacks.



\vspace{-0.5mm}
\noindent{\bf Resiliency to model modifications (Robustness requirement).}
We analyze the robustness of DynaMarks scheme against the following types of watermark removal strategies. 



\vspace{-1mm}
\noindent{\bf (i)} {\em Averaging Attack.}
If an attacker queries $\mathcal{F}_{alt}$ repeatedly using the same input, then according to Algorithm~\ref{algo_wm_embed} she will get different output probabilities for different input queries as the model response is dependent on a couple of randomness factors (lines $6$ and $7$ of Algorithm~\ref{algo_wm_embed}).  
The attacker can then calculate the average of such output probabilities to get a representative response for an input which are used to populate the substitute dataset $D_{sub}$ for training $\mathcal{F}_{sm}$.
The attacker succeeds if the obtained surrogate model $\mathcal{F}_{sm}$ does not retain any information pertaining to the watermark.
In order to evaluate the outcome of such an attack for a data fraction $\gamma$ = $1$ (entire training data), we queried $\mathcal{F}_{alt}$ repeatedly using the same input for $100$ times to get a representative sample in the substitute dataset $D_{sub}$.
Then, we obtained the distance metrics $\delta^{org}_{sm}$ and $\delta^{alt}_{sm}$ as reported in subcolumns $4$ and $5$ of Table~\ref{table:DynaMarks_evaluation}.
We observe that their ratio $\eta$ (see subcolumn $6$) is still greater than the watermark detection threshold $\tau$ = $1$, implying that DynaMarks is resistant to such averaging attack.


\vspace{-1mm}
\noindent{\bf (ii)} {\em Model Compression.}
The attacker can also adopt a model pruning approach to compress the surrogate model $\mathcal{F}_{sm}$ with the objective of removing the watermark~\cite{jia2020entangled,rouhani2018deepsigns}.
In our experiments, to perform model compression, we eliminated the lowest $\kappa \%$ of the network connections across all the layers of $\mathcal{F}_{sm}$ (trained using data fraction $\gamma$ = $1$) using the global pruning method of PyTorch framework.
We report the outcomes of such model pruning with a pruning rate $\kappa$ = $10\%$
in subcolumns $7$-$9$ of Table~\ref{table:DynaMarks_evaluation}.
In this case also we observe that the watermark detection parameter $\eta$ is above the threshold $\tau$ = $1$ for all the three datasets, highlighting the robustness of DynaMarks scheme against model compression.
We further varied the pruning rate $\kappa$ from $10\%$ up to $50\%$; we find in Fig.~\ref{fig_impact_pruning_rate} that even then the watermark can be successfully detected ($\eta > \tau$) across all the pruning rates.
It is to be noted that high proportions of pruning result in significant accuracy degradation of the extracted model $\mathcal{F}_{sm}$ with negligible impact on the detection parameter $\eta$, e.g., in case of Fashion MNIST dataset the model accuracy drops by $11.17\%$ with $\kappa$ = $50\%$ compared to the uncompressed model accuracy even though for the pruned model the parameter $\eta$ = $2.11$ remains well above the threshold $\tau$ = $1$.

\vspace{-1mm}
\noindent{\bf (iii)} {\em Different architectures.}
All the previous set of experiments were performed by keeping the same network architecture for the victim and the surrogate models.
In order to study the influence of the choice of architecture on DynaMarks technique:
(a) in case of Fashion MNIST and CIFAR-10 datasets, we use ResNet-18 network~\cite{he2016deep} for obtaining a surrogate model $\mathcal{F}_{sm}$ from a CNN-based victim model $\mathcal{F}_{alt}$ with a data fraction $\gamma$ = $1$. 
(b) in case of ImageNet dataset, we use VGG-11~\cite{simonyan2014very} for obtaining $\mathcal{F}_{sm}$ from a ResNet-18-based $\mathcal{F}_{alt}$ with $\gamma$ = $1$.
From the last three subcolumns of Table~\ref{table:DynaMarks_evaluation}, we observe that even in this setting the watermark detection parameter $\eta$ is sufficiently greater than the threshold $\tau$ = $1$ for all the image datasets considered.
This highlights the fact that the watermark generated by altering the responses of the victim DL model using Algorithm~\ref{algo_wm_embed} transfers to the responses of the surrogate model $\mathcal{F}_{sm}$ irrespective of its architectural choice.
This is because $\mathcal{F}_{sm}$ replicates the input-output mapping of the model $\mathcal{F}_{alt}$ (which is a function of the secret parameters chosen by the DL model owner) provided that the network architecture of $\mathcal{F}_{sm}$ is sufficiently complex.



\vspace{-0.5mm}
\noindent{\bf Imperceptibility requirement.}
Unlike prior black-box watermarking approaches~\cite{jia2020entangled, szyller2019dawn}, DynaMarks does not utilize any trigger inputs to detect watermark in a surrogate model $\mathcal{F}_{sm}$ obtained using model extraction attack.
Hence, state-of-the-art backdoor detection schemes such as Neural Cleanse~\cite{wang2019neural} are not applicable to DynaMarks which generates watermark by dynamically altering the responses of the prediction API of a protected DL model at inference runtime.
Also, it seems very unlikely that a data-limited attacker will be able to revert back such alterations without any notion of the secret parameters used in the watermark generation process.
As future work, we plan to investigate the security offered by DynaMarks technique against steganalysis and watermark overwriting attacks.



\vspace{-2.0mm}
\section{Conclusion}
\label{sec:DynaMarks_concl}
\vspace{-1mm}

In this paper, we present a novel watermarking approach called DynaMarks as an effective IP security solution against model extraction attacks on DL models deployed in edge devices. 
Unlike existing defenses, our proposed scheme does not introduce any computational overhead in the model training phase nor does it sacrifice the victim model's prediction accuracy
to gain security benefits.
DynaMarks embeds watermark into a surrogate model by dynamically changing the responses of the victim model's prediction API based on certain secret parameters at inference runtime. 
The experimental outcomes on Fashion MNIST, CIFAR-10, and ImageNet datasets demonstrate the effectiveness and robustness of DynaMarks technique against different types of watermark removal strategies.

\bibliographystyle{abbrv}
{
\bibliography{bib_file.bib,ref-yxie.bib} 

\begin{thebibliography}{10}

\bibitem{cifar10_cite}
{CIFAR-10 dataset}.
\newblock https://www.cs.toronto.edu/~kriz/cifar.html.

\bibitem{fashion_mnist_cite}
{Fashion MNIST}.
\newblock https://github.com/zalandoresearch/fashion-mnist.

\bibitem{cite_tpm}
{TPM}.
\newblock \url{https://trustedcomputinggroup.org/}.

\bibitem{adi2018turning}
Y.~Adi~et al.
\newblock Turning your weakness into a strength: Watermarking deep neural
  networks by backdooring.
\newblock In {\em USENIX}, pages 1615--1631, 2018.

\bibitem{atli2019extraction}
B.~G. Atli~et al.
\newblock Extraction of complex dnn models: Real threat or boogeyman?
\newblock {\em arXiv preprint arXiv:1910.05429}, 2019.

\bibitem{chakraborty2020hardware}
A.~Chakraborty~et al.
\newblock Hardware-assisted intellectual property protection of deep learning
  models.
\newblock In {\em 2020 57th ACM/IEEE Design Automation Conference (DAC)}, pages
  1--6. IEEE, 2020.

\bibitem{goodfellow2016deep}
I.~Goodfellow~et al.
\newblock {\em Deep Learning}.
\newblock MIT press, 2016.

\bibitem{gu2017badnets}
T.~Gu~et al.
\newblock Badnets: Identifying vulnerabilities in the machine learning model
  supply chain.
\newblock {\em arXiv preprint arXiv:1708.06733}, 2017.

\bibitem{guo2018watermarking}
J.~Guo~et al.
\newblock Watermarking deep neural networks for embedded systems.
\newblock In {\em 2018 IEEE/ACM International Conference on Computer-Aided
  Design (ICCAD)}, pages 1--8. IEEE, 2018.

\bibitem{he2016deep}
K.~He~et al.
\newblock Deep residual learning for image recognition.
\newblock In {\em Conference on computer vision $\&$ pattern recognition},
  pages 770--778. IEEE, 2016.

\bibitem{howard2020fastai}
J.~Howard and S.~Gugger.
\newblock Fastai: a layered api for deep learning.
\newblock {\em Information}, 11(2):108, 2020.

\bibitem{jia2020entangled}
H.~Jia~et al.
\newblock Entangled watermarks as a defense against model extraction.
\newblock In {\em 30th USENIX Security Symposium (USENIX Security 21)}, pages
  1937--1954, 2021.

\bibitem{kariyappa2020defending}
S.~Kariyappa~et al.
\newblock Defending against model stealing attacks with adaptive
  misinformation.
\newblock In {\em Proceedings of the IEEE/CVF Conference on Computer Vision and
  Pattern Recognition}, pages 770--778, 2020.

\bibitem{lee2018defending}
T.~Lee~et al.
\newblock Defending against machine learning model stealing attacks using
  deceptive perturbations.
\newblock {\em arXiv preprint arXiv:1806.00054}, 2018.

\bibitem{li2018edge}
E.~Li~et al.
\newblock Edge intelligence: On-demand deep learning model co-inference with
  device-edge synergy.
\newblock In {\em Proceedings of the 2018 Workshop on Mobile Edge
  Communications}, pages 31--36, 2018.

\bibitem{lukas2019deep}
N.~Lukas~et al.
\newblock Deep neural network fingerprinting by conferrable adversarial
  examples.
\newblock {\em arXiv preprint arXiv:1912.00888}, 2019.

\bibitem{merrer2017adversarial}
E.~Merrer~et al.
\newblock Adversarial frontier stitching for remote neural network
  watermarking.
\newblock {\em arXiv preprint arXiv:1711.01894}, 2017.

\bibitem{nagai2018digital}
Y.~Nagai, Y.~Uchida, S.~Sakazawa, and S.~Satoh.
\newblock Digital watermarking for deep neural networks.
\newblock {\em International Journal of Multimedia Information Retrieval},
  7(1):3--16, 2018.

\bibitem{orekondy2019knockoff}
T.~Orekondy~et al.
\newblock Knockoff nets: Stealing functionality of black-box models.
\newblock In {\em Proceedings of the IEEE Conference on Computer Vision and
  Pattern Recognition}, pages 4954--4963, 2019.

\bibitem{orekondy2019prediction}
T.~Orekondy~et al.
\newblock Prediction poisoning: Utility-constrained defenses against model
  stealing attacks.
\newblock {\em arXiv preprint arXiv:1906.10908}, 2019.

\bibitem{pal2019framework}
S.~Pal~et al.
\newblock A framework for the extraction of deep neural networks by leveraging
  public data.
\newblock {\em arXiv preprint arXiv:1905.09165}, 2019.

\bibitem{papernot2017practical}
N.~Papernot~et al.
\newblock Practical black-box attacks against machine learning.
\newblock In {\em Proceedings of the 2017 ACM on Asia CCS}, pages 506--519,
  2017.

\bibitem{rouhani2018deepsigns}
B.~D. Rouhani~et al.
\newblock Deepsigns: An end-to-end watermarking framework for ownership
  protection of deep neural networks.
\newblock In {\em Proceedings of International Conference on ASPLOS}, pages
  485--497. ACM, 2019.

\bibitem{simonyan2014very}
K.~Simonyan and A.~Zisserman.
\newblock Very deep convolutional networks for large-scale image recognition.
\newblock {\em arXiv preprint arXiv:1409.1556}, 2014.

\bibitem{szyller2019dawn}
S.~Szyller~et al.
\newblock Dawn: Dynamic adversarial watermarking of neural networks.
\newblock {\em arXiv preprint arXiv:1906.00830}, 2019.

\bibitem{tramer2016stealing}
F.~Tram{\`e}r~et al.
\newblock Stealing machine learning models via prediction apis.
\newblock In {\em 25th USENIX Security Symposium}, pages 601--618, 2016.

\bibitem{uchida2017embedding}
Y.~Uchida~et al.
\newblock Embedding watermarks into deep neural networks.
\newblock In {\em Proceedings of the 2017 ACM on International Conference on
  Multimedia Retrieval}, pages 269--277. ACM, 2017.

\bibitem{wang2019neural}
B.~Wang~et al.
\newblock Neural cleanse: Identifying and mitigating backdoor attacks in neural
  networks.
\newblock In {\em 2019 IEEE Symposium on Security and Privacy (SP)}, pages
  707--723. IEEE, 2019.

\bibitem{wang2019robust}
T.~Wang~et al.
\newblock Robust and undetectable white-box watermarks for deep neural
  networks.
\newblock {\em arXiv preprint arXiv:1910.14268}, 2019.

\bibitem{wu2019machine}
C.-J. Wu~et al.
\newblock Machine learning at facebook: Understanding inference at the edge.
\newblock In {\em 2019 IEEE International Symposium on High Performance
  Computer Architecture (HPCA)}, pages 331--344. IEEE, 2019.

\bibitem{zhang2018protecting}
J.~Zhang~et al.
\newblock Protecting intellectual property of deep neural networks with
  watermarking.
\newblock In {\em Proceedings of Asia CCS}, pages 159--172, 2018.

\end{thebibliography}
}

\end{document}